
%

\magnification \magstep1

\hoffset 1.5truecm
\hsize 16truecm \vsize 23truecm
\baselineskip 20pt \parskip5pt
\raggedbottom

\font\cs=cmcsc10

\def\intpar{\hfil\break\indent}
\def\dim{{\rm dim}}
\def\mod{{\rm mod}\,}

\def\clap#1#2{
\setbox0=\hbox{$\displaystyle#2$}
\hbox to \wd0 {\hss$#1$\hss}
\kern-\wd0\hbox to\wd0{\hss$\displaystyle#2$\hss}}


\rightline{hep-th/9305101}

\vfill

\centerline
{\bf A note on the four-dimensional Kirby calculus}

\bigskip

\centerline{BOGUS\L AW BRODA}

\vfill

\noindent
{\cs Abstract.}
An explicit derivation of the elements of the
representation ring of ${\rm SU}(2)$ needed to implement
the four-dimensional Kirby calculus is sketched.

\vfill

\centerline{\cs May 1993}

\vfill

\footnote{}%
{1991 {\it Mathematics Subject Classification}.  Primary
57N13; Secondary 57M25,
\hfil\break
57N10, 57R65.
\intpar
}

\vfill\eject

The aim of this short note is to explicitly derive, in an
elementary way, the elements $\Omega$, $\omega$ of the
representation ring of the group ${\rm SU}(2)$ needed to
implement the four-dimensional Kirby calculus (handle
slides) of C\'esar~de~S\'a [1], used in the construction of
the surgical invariant of ref.~[2].

Sliding an upper curve over a lower one is schematically
depicted as
$$
{\bigcup_\Omega \atop \bigodot_\omega}
\longleftrightarrow
\clap{\odot}{\bigcup^\omega}_\Omega,
\eqno{(1)}
$$
where
$$
\Omega=\delta_m W_m,
\qquad
\omega=\bigoplus_{n=0}^N \Delta_n W_n,
\eqno{(2)}
$$
with $W_n$ an irreducible module of ${\rm SU}(2)$,
$\dim\,W_n=n+1$ (we use the terminology of ref.~[3]).

Applying the formula for a connected sum, as well as the
satellite formula [3], we obtain, for each $m$, the
fundamental relation [4] (see also [5])
$$
\dim_q (W_m) \Delta_n
=\sum_k C_{k m}^n \Delta_k,
\qquad
n:\dim_q W_n \not= 0,
\eqno{(3)}
$$
where $C_{km}^n$ are the classical Clebsch-Gordan
coefficients, and the quantum dimension
$$
\dim_q W_m
=(-)^m {q^{m+1\over2} - q^{-{m+1\over2}}
\over q^{1\over2} - q^{-{1\over2}}}.
\eqno{(4)}
$$
The deformation parameter
$$
q=e^{2\pi i\over r},
\eqno{(5)}
$$
where $r$ is the level, a positive integer.

\bigskip

Thus, we have the following cases:

\noindent
\underbar{$m=1$} ($r\geq2$)

Explicitly, eq.~(3) is of the form
$$
-\left(q^{1\over2}+q^{-{1\over2}}\right) \Delta_n
=\Delta_{n+1} + \Delta_{n-1},
\qquad
n \not= -1\ \mod r,
\eqno{(6)}
$$
with the initial conditions
$$
\Delta_{-1}=0,
\qquad
\Delta_0=a_0,
\eqno{(7)}
$$
$a_0$ an arbitrary complex number. The only non-trivial
solution is given by
$$
\Delta_n
=\cases{a_0\dim_q W_n,&if $0\leq n\leq r-2$;\cr
0,&otherwise.\cr}
\eqno{(8)}
$$

\noindent
\underbar{$m=2$} ($r\geq3$)

Now, eq.~(3) takes the form
$$
\left(q+q^{-1}\right) \Delta_n
=\Delta_{n+2} + \Delta_{n-2},
\qquad
n\not=-1\ \mod r,
\eqno{(9)}
$$
and decouples for even and odd $n$. Assuming the initial
conditions
$$
\Delta_{-2}=\Delta_{-1}=0,
\qquad
\Delta_0=a_+,
\qquad
\Delta_1=a_-\dim_q W_1,
\eqno{(10)}
$$
where $a_+$, $a_-$ are arbitrary complex numbers, we obtain
$$
\Delta_n
=\cases{
a_+\dim_q W_n, & for $r$ odd, if $n$ even and $0\leq n\leq
r-3$;\cr
a_-\dim_q W_n, & for $r$ even, if $n$ odd and $1\leq n\leq
r-3$;\cr
0, & otherwise.\cr}
\eqno{(11)}
$$

\noindent
\underbar{$n\raise.3ex\hbox{$\geq$}3$}

There are no non-trivial, i.\thinspace e.\ non-zero with a
finite $N$, solutions of eq.~(3).

\bigskip

Taking into account that any $W_n$ can be built of $W_1$,
and any $W_n$ with $n$ even can be built of $W_2$, we
obtain for $r$ odd the desired moves
$$
{\bigcup_{\omega_0} \atop \bigodot_{\omega_0}}
\longleftrightarrow
\clap{\odot}{\bigcup^{\omega_0}}_{\omega_0},
\qquad
{\bigcup_{\omega_+} \atop \bigodot_{\omega_0}}
\longleftrightarrow
\clap{\odot}{\bigcup^{\omega_0}}_{\omega_+},
\qquad
{\bigcup_{\omega_+} \atop \bigodot_{\omega_+}}
\longleftrightarrow
\clap{\odot}{\bigcup^{\omega_+}}_{\omega_+},
\eqno{\rm(12a)}
$$
but
$$
{\bigcup_{\omega_0} \atop \bigodot_{\omega_+}}
\clap{/}{\longleftrightarrow}
\clap{\odot}{\bigcup^{\omega_+}}_{\omega_0},
\eqno{\rm(12b)}
$$
where
$$
\omega_0=\bigoplus_{n=0}^{r-2} a_0 \dim_q (W_n) W_n,
\qquad
\omega_+=\bigoplus_{n=0\atop n\ {\rm even}}^{r-3} a_+ \dim_q
(W_n) W_n.
\eqno{(13)}
$$

\bigskip
\bigskip

The author is indebted to Prof.  H.~D.~Doebner for his kind
hospitality in Clausthal. The work was partially  supported
by the CEC grant CIPA3510PL921596, the KBN grant 202189101,
and the University of \L\'od\'z grant.

\vfill\eject

\noindent
{\cs References}

\bigskip

{\frenchspacing

\item{1.} E. C\'esar de S\'a, {\it A link calculus for
4-manifolds}, Topology of low-dimensional manifolds, Proc.
Second Sussex Conf., Lecture Notes in Math., vol. {\bf
722}, Springer, Berlin, 1979, 16--30.

\item{2.} B. Broda, {\it Surgical invariants of
four-manifolds}, E-preprint hep-th/9302092.

\item{3.} H. R. Morton and P. Strickland, Math. Proc. Camb.
Phil. Soc. {\bf 109} (1991) 83--103.

\item{4.} B. Broda, {\it Chern-Simons theory on an
arbitrary manifold via surgery}, E-preprint hep-th/9305051.

\item{5.} K. Walker, {\it On Witten's 3-manifold
Invariants}, preprint (1990).

\bigskip
\bigskip}

{\cs
Department of Theoretical Physics, University of \L\'od\'z,
Pomorska 149/153, PL--90-236 \L\'od\'z, Poland
}

{\it Current address\/}:
Arnold Sommerfeld Institute for Mathematical Physics,
Technical University of Clausthal, Leibnizstra\ss e 10,
D-W--3392 Clausthal-Zellerfeld, Federal Republic of Germany

{\it E-mail address\/}:
ptbb@ibm.rz.tu-clausthal.de

\bye